\documentclass[prd,showpacs,amsmath,amssymb,11pt]{revtex4}
\usepackage{latexsym}
\usepackage{bm}

\begin{document}

\title{Finite mass gravitating Yang monopoles}

\author{Hakan Cebeci}
\email{hcebeci@anadolu.edu.tr}
\affiliation{Department of Physics, Anadolu University, 26470, Eski{\c s}ehir, Turkey}

\author{{\" O}zg{\" u}r Sar{\i}o\u{g}lu}
\email{sarioglu@metu.edu.tr}
\affiliation{Department of Physics, Faculty of Arts and  Sciences,\\
             Middle East Technical University, 06531, Ankara, Turkey}

\author{Bayram Tekin}
\email{btekin@metu.edu.tr}
\affiliation{Department of Physics, Faculty of Arts and  Sciences,\\
             Middle East Technical University, 06531, Ankara, Turkey}

\date{\today}

\begin{abstract}
We show that gravity cures the infra-red divergence of the Yang monopole 
when a proper definition of conserved quantities in curved backgrounds is used, 
i.e. the gravitating Yang monopole in cosmological Einstein theory has a 
finite mass in generic even dimensions (including time). In addition, we find 
exact Yang-monopole type solutions in the cosmological Einstein-Gauss-Bonnet-Yang-Mills 
theory and briefly discuss their properties.
\end{abstract}

\pacs{11.15.-q, 04.50.Gh, 14.80.Hv, 04.50.Kd}

\maketitle

\section{\label{intro} Introduction}

About 30 years ago, Yang \cite{yang}, generalizing the Dirac monopole, found a 
(singular) spherically symmetric solution of the five-dimensional Euclidean 
Yang-Mills (YM) theory with the $SU(2)$ gauge group. In the same paper, he also
showed that his $SU(2)$ monopole does not exist in more than five dimensions.
Yang's monopole (on which we shall dwell a bit more in a moment) went pretty 
much unnoticed up until it emerged in a rather unlikely place, in the study 
of the four-dimensional analog of the quantum Hall effect \cite{zhang}. [We 
have nothing more to say about the Yang monopole in its relevance to the 
quantum Hall effect, except to remark that no solution of the YM theory seems 
to be wasted!]

The present work was inspired by and follows closely the recent article by
Gibbons and Townsend \cite{gib}, which does a couple of things at once. It
introduces gravity into the picture to get gravitating Yang monopoles, and
gives a reconstruction (and reinterpretation) of the higher dimensional 
versions (with gauge groups other than $SU(2)$) of both the curved and the 
flat space Yang monopoles. [See \cite{hp} for an earlier discussion of the
higher dimensional Yang monopoles.] Before we explain how we ``improve'' on 
the work of Gibbons-Townsend, let us recapitulate some properties of the Yang 
monopole.

The way Yang constructed his solution is quite interesting: He considered 
self-dual, spherically symmetric single instanton (and anti-instanton) solutions 
on $S^4$ and showed that they solve the full YM equations in five Euclidean 
dimensions. As five-dimensional solutions, these instantons have a singularity at 
the origin just like their three-dimensional cousin, ``the Dirac monopole''. The 
action of the single self-dual instanton, \( \int {\cal F} \wedge {\cal F} \) 
integrated over $S^{4}$, becomes a conserved monopole charge of the 
five-dimensional Yang monopole. [Note that even though there are instantons whose 
charge can take an arbitrary integral value in four dimensions, none save the $\pm 1$ 
charge solves the five-dimensional YM equations. Put in another way, there are no 
Yang multi-monopoles! This is a curious result, but can be shown to be valid by 
topological arguments \cite{hp}, as we will also argue.] As summarized in \cite{gib}, 
the rare appearance of the Yang monopole in high energy physics literature might be 
due to the fact that unlike the ultra-violet (UV) divergence
[\( \int d^{3}x B^{2} \to g^{2} \int_{0}^{\infty} dr/r^{2} \to \infty\)] of the 
Dirac monopole, the Yang monopole has an infra-red (IR) divergence, i.e. its mass 
is IR divergent. We know that when compact Maxwell theory with Dirac monopoles is 
considered as a low energy limit of, say, a broken $SO(3)$ Georgi-Glashow type theory, 
finite mass 't Hooft-Polyakov monopoles emerge, which look exactly like Dirac monopoles 
from a distance. Therefore, UV divergence of the Dirac monopole is not a great concern 
if some unified theory picture is adopted. In the case of the Yang monopole, one 
needs to construct a microscopic theory which takes over in the IR limit, which, 
of course, is quite a difficult task. [See \cite{rt, khn} where some higher derivative 
YM actions with Higgs fields are used to construct regular monopole solutions in 
higher dimensions.]

Note that all of the discussion about the mass-divergence of the Yang monopole above 
is in flat space. If we turn on gravity, as we shall do in this paper, the picture 
changes drastically. Gravity could be blamed for introducing UV divergences, curvature 
singularities and black holes, but since it clumps matter and fields, it should be a 
good cure for IR divergences. Gibbons-Townsend \cite{gib} introduced the self-gravitating 
Yang monopole and argued that, in contrast to this expectation, the mass is still IR 
divergent (beyond four dimensions in their classification). Here, we show that once 
the proper mass-energy definitions in asymptotically flat and AdS spaces are employed, 
the Yang monopole does indeed have a finite mass in all dimensions. The main issue here 
has to do with the choice of a proper background to work out the relevant mass-energy 
formula.

Our second aim in this paper is to find Yang-monopole type solutions in more generic 
gravity theories coupled with YM systems. To this end, we consider the cosmological 
Einstein-Gauss-Bonnet (GB) theory, which appears as a low energy limit of some string 
theories, and construct new solutions. Compared to General Relativity, GB theory behaves 
better in the UV region, which is not our main concern here, but exact solutions in this 
rather complicated theory are always good to have, since there are very few known anyway.

The organization of this paper is as follows. In the next section, we briefly review the 
Dirac and Yang monopoles in flat space. In section \ref{efmo}, we show how the IR
divergence mentioned above is overcome. We describe the cosmological Einstein-GB-YM theory
in section \ref{set}, and present our ansatz for the Yang monopole, our assumptions
and the field equations we obtained from these in section \ref{eqns}. Section \ref{solns}
is devoted to the solutions found and their properties. Finally we conclude with section 
\ref{conc}.

\section{\label{mono} Dirac and Yang monopoles in flat space}

As there can be occasional confusions with regard to gauge symmetries, spacetime 
symmetries and the charge definitions of higher dimensional singular monopoles, we 
start by giving a brief recollection of these concepts in flat space, with the help
of \cite{eh} and \cite{naka}. 

Let us start with Yang's generalization of the three-dimensional Dirac monopole. 
The latter lives on $\mathbb{R}^{3}$ with the origin removed. The Maxwell field 
strength ${\cal F}$ is a 2-form whose flux $\int_{S^{2}} {\cal F}$ gives the magnetic 
charge which can take \emph{any} integral value up to a normalization. Even though 
the vector potential ${\cal A}$ does not reflect it, the physical field ${\cal F}$ 
is spherically symmetric, i.e. it is invariant under the action of $SO(3)$. [This in
fact means that spatial rotations can be undone with gauge transformations.] As is 
well known, the singular Dirac monopole can be described by pure geometry: 
$\mathbb{R}^{3}-\{0\}$ is homotopically equivalent to $S^{2}$, therefore one may study 
the corresponding principal bundle $P(S^{2},U(1))$. [For the charge-1 monopole, this 
is the Hopf fibration of $S^{3}$.] Then the 
transition functions defined on the equator $S^{1}$ of $S^{2}$ classify the magnetic 
charge; namely, they map $S^{1} \to U(1)$, having $\pi_{1}(U(1)) = \mathbb{Z}$. A 
complementary picture is provided by the first Chern character of this ``monopole'' 
bundle, i.e. the magnetic flux equals $\int_{S^{2}} ch_{1}({\cal F})$.

Let us now look at the ``original'' Yang monopole \cite{yang} in $\mathbb{R}^{5}-\{0\}$, 
which is homotopically equivalent to $S^{4}$. Yang considered the field strength 
${\cal F}$ to be an $\mathfrak{su}(2)$-valued 2-form that ``generalizes'' the Dirac 
monopole in the sense that the physically measurable quantities are $SO(5)$ invariant. 
Now the relevant geometrical object is the principal bundle $P(S^{4},SU(2))$. Even though 
the corresponding homotopy group $\pi_{3}(SU(2))$ equals $\mathbb{Z}$ which arises from 
the maps $S^{3} \to SU(2)$, the Euclidean YM equation in five dimensions admits only 
\emph{two} of these solutions. These are just the four-dimensional self-dual and anti 
self-dual solutions (BPST instanton having the charge $\pm 1$). The charge is now given 
(up to a normalization) by the integral of the second Chern character 
\[ \int_{S^{4}} ch_{2}({\cal F}) = \int_{S^{4}} \, \mbox{Tr} \, ({\cal F} \wedge {\cal F}) = \pm 1 \, . \]

\section{\label{efmo} The effect of gravity on monopoles}

Here we will show how gravity cures the IR divergence of the mass-energy of the Yang monopole 
in any even dimensions (time included), just as it cures the UV divergence of the 3+1-dimensional 
Dirac monopole. [In this context, the latter is nothing but the celebrated Reissner-Nordstrom 
black hole.] Note that our result about the mass of the Yang monopole is not in agreement with 
\cite{gib}, who incorrectly claimed that the divergence persisted in the presence of gravity 
except for four dimensions. The gist of the problem lies in the correct definition of gravitational 
mass-energy. For this purpose, we resort to the procedure developed in \cite{ad, dt1, dt2}. 
Stated briefly, the idea is to define gauge invariant conserved charges in a diffeomorphism 
invariant theory by employing the generalized ``Gauss law'' provided there exist asymptotic 
Killing symmetries of the relevant spacetimes. Put in another way, one chooses a vacuum that 
satisfies the field equations as the background with respect to which background gauge 
invariant quantities (such as energy) is calculated. These charges are expressible as 
surface integrals and, by construction, their value for the background itself is always 
zero. The latter is quite important.

Given a background Killing vector $\bar{\xi}^{\mu}$, the corresponding conserved charges
can be written as \footnote{Throughout, we set the Newton's constant $G_{n} = 1$.} \cite{dt1, dt2}
\begin{equation} 
Q^{\mu}(\bar{\xi}) = \frac{1}{4 \, \Omega_{n-2}} \,
\int d^{n-2} x \, \bar{\xi}_{\nu} \, {\cal G}^{\mu\nu}_{L} \,, \label{einyuk}
\end{equation}
where ${\cal G}^{\mu\nu}_{L}$ denotes the linearized Einstein tensor about 
the background and $\Omega_{n-2}$ is the solid angle on the unit $(n-2)$-sphere. 
As it would be too much of a digression to redrive this formula and its form 
as a surface integral, we refer the reader to \cite{dt1, dt2} for the details 
and simply employ it here.

For the gravitating Yang-monopole type solutions found in \cite{gib}, the 
spacetime metric in Schwarzschild-like coordinates is simply given by
\begin{equation} 
ds^{2} = - f^{2}(r) \, dt^{2} + \frac{dr^{2}}{f^{2}(r)} + r^{2} \,
d \Omega^{2}_{n-2} \,, \label{gibsol}
\end{equation} 
where $d \Omega^{2}_{n-2}$ is the metric on the $(n-2)$-sphere and the function 
$f(r)$ reads (in the form presented by \cite{gib} but adapted to our conventions 
for $n \geq 4$)
\begin{equation}
f^{2}(r) = 1 - \frac{2 \, m}{r^{n-3}} - \frac{\mu^{2}}{r^{2}} -
\frac{2 \, \Lambda \, r^{2}}{(n-2)(n-1)} \,. \label{fgib}
\end{equation}
Here the constant $\mu$ is given by
\[ \mu^{2} = \frac{8 \pi (n-3)}{(n-5) \sigma^{2}} \,, \] 
which follows from the normalization choice for the generators $\Sigma_{ij}$ 
of the gauge group $SO(n-2)$ (see \cite{gib} for details), and cannot be chosen 
as zero. This is a rather important point. Together with the cosmological term, the
$\mu^{2}$ piece in (\ref{fgib}) constitute the background with respect to which
any spacetime with nonvanishing $m$ can have a finite and meaningful mass. 
Otherwise, apart from the special $n=4$ case, one always finds a divergent mass for 
(\ref{gibsol}). Thus taking the background to be the spacetime (\ref{gibsol}) 
with $m=0$ in (\ref{fgib}), which has the timelike Killing vector 
\( \bar{\xi}^{\mu} = (-1, 0, \dots, 0) \) in the notation and conventions 
of \cite{dt1, dt2}, one finds the total energy of these solutions as
\[ E = \frac{1}{4 \Omega_{n-2}} \, \Omega_{n-2} \, (2(n-2) \, m)
= \frac{m(n-2)}{2} \,. \]
This is the result of the surface integration at $r \to \infty$ in the notation of 
\cite{dt1, dt2}. To see how gravity modifies the IR divergence of the Yang monopole,
let us also compute the (gauge non-invariant) energy contained in a ball of radius
$R$ about the origin of spacetime. One then finds
\begin{equation}
E(R) = \frac{(n-2) \, m \, R^{n-5} \, \left[ 2 \, \Lambda \, R^{4} + (n-1)(n-2) (\mu^{2} - R^{2}) \right]}
{2 \, \big( \left[ 2 \, \Lambda \, R^{4} + (n-1)(n-2) (\mu^{2} - R^{2}) \right] R^{n-5} + 2 (n-1)(n-2) \, m \big)} \,,
\end{equation}
which is finite in contrast to the flat space result, that goes like $R^{n-5}$ and
diverges as $R \to \infty$ for $n \geq 6$ \cite{gib}.

\section{\label{set} The cosmological Einstein-GB-YM theory}

Let us now describe the cosmological Einstein-GB-YM theory, the assumptions we make 
and the solutions they lead to in various dimensions. We start with the action
\begin{equation} 
I[e, {\cal A}] = \int {\cal L} \,, \label{act}
\end{equation}
where the Lagrangian density $n$-form 
\begin{eqnarray}
{\cal L} & = & \frac{1}{2}\, R^{ab} \wedge \ast (e_{a} \wedge e_{b}) 
- \frac{1}{2 \sigma^{2}}\, \mbox{Tr} ({\cal F} \wedge \ast {\cal F} ) + \Lambda \ast 1 
+ \frac{\gamma}{4}\, R^{ab} \wedge R^{cd} \wedge \ast ( e_{a}
\wedge e_{b} \wedge e_{c} \wedge e_{d} ) \label{lag}
\end{eqnarray}
contains the Einstein-Hilbert term, the YM Lagrangian for the 2-form
field ${\cal F}$ with coupling constant $\sigma$, a cosmological constant $\Lambda$ and
a second order Euler-Poincar\'{e} term (the so called GB term in this case) with coupling 
constant $\gamma$. 

The basic gravitational field variables are the coframe 1-forms $e^{a}$ in terms 
of which the spacetime metric is decomposed as 
\( {\mbox{\bf g}} = \eta_{ab} \, e^{a} \otimes e^{b} \), where 
\( \eta_{ab} = \mbox{diag} \, ( -, +, +, \dots ) \) is the Minkowski metric. 
The Hodge duality map is specified by the oriented volume element
\( \ast 1 = e^{0} \wedge e^{1} \wedge \dots \wedge e^{n-2} \wedge e^{n} \). The 
torsion-free, Levi-Civita connection 1-forms $\omega^{a}\,_{b}$ satisfy the first 
Cartan structure equations
\[ d e^{a} + \omega^{a}\,_{b} \wedge e^{b} = 0, \]
where metric compatibility implies \( \omega_{ab} = - \omega_{ba} \). The corresponding 
curvature 2-forms follow from the second Cartan structure equations
\[ R^{a}\,_{b} = d \omega^{a}\,_{b} + \omega^{a}\,_{c} \wedge \omega^{c}\,_{b} \,. \] 
The GB term in the Lagrangian density (\ref{lag}) can also be written in the alternative form 
\[ R^{ab} \wedge R^{cd} \wedge \ast ( e_{a} \wedge e_{b} \wedge e_{c} \wedge e_{d} ) =
2 \, R_{ab} \wedge \ast R^{ab} - 4 \, P_{a} \wedge \ast P^{a} + {\cal R}_{(n)}^{2} \ast 1 \,, \]
where the Ricci 1-form \( P^{a} = \iota_{b} R^{ba} \) and the curvature scalar 
\( {\cal R}_{(n)} = \iota_{a} \, \iota_{b} \, R^{ba} \) have been utilized via the interior
product operator \( \iota_{a} = \iota_{X^{a}} \) for which \( \iota_{X^{b}} (e^a) = \delta_b\,^{a} \).

Before moving on to the field equations, let us present our setting on the YM sector
as well. We take the YM potential ${\cal A}$ to be a Lie algebra $\mathfrak{g}$-valued 1-form.
The YM 2-form field follows from 
\begin{equation} 
{\cal F} = d {\cal A} + \frac{1}{2} \, [{\cal A}, {\cal A}] \label{ym2f}
\end{equation}
in the usual way and satisfies the Bianchi identity
\begin{equation} 
D {\cal F} = d {\cal F} + [{\cal A}, {\cal F}] = 0 \,. \label{bian} 
\end{equation}

The field equations read 
\begin{eqnarray}
\frac{1}{2} \, R^{ab} \wedge \ast ( e_{a} \wedge e_{b} \wedge e_{c} )
& = & - \frac{1}{4 \sigma^{2}} \tau_{c} [{\cal F}] - \Lambda \ast e_{c} 
- \frac{\gamma}{4} \, R^{ab} \wedge R^{dg} \wedge \ast ( e_{a} \wedge
e_{b} \wedge e_{d} \wedge e_{g} \wedge e_{c} ) \,, \label{eingb} \\
D \ast {\cal F} & = & d \ast {\cal F} + [{\cal A}, \ast {\cal F} ] = 0 \,. \label{ymeq}
\end{eqnarray}
Here
\begin{equation}
\tau_{c} [{\cal F}] = 2 \, \mbox{Tr} \, \left( \iota_{c} {\cal F} \wedge \ast {\cal F} 
- {\cal F} \wedge \iota_{c} \ast {\cal F} \right) \label{emten}
\end{equation}
is the corresponding stress-energy $(n-1)$-form for the gauge field ${\cal F}$.

\section{\label{eqns} The Ans{\"a}tze and equations for the fields}

Following \cite{yang} and \cite{gib}, we will consider solutions that have field 
strengths only on an $(n-2)$-sphere. [Namely, there will be no radial components. 
In fact, as explained in \cite{gib}, when radial components are introduced, one 
usually gets a different class of (numerical) solutions such as the ones obtained by 
Bartnik-McKinnon in four dimensions \cite{bm}.] This naturally leads to the choice of 
the gauge group $G$ to be $SO(n-2)$ (for $n \geq 4$) and the Ans{\"a}tze for the 
metric and gauge potential follow accordingly. Let us decompose the local coordinates 
for the spacetime as
\[ x^{M} = \left\{ x^{0} \equiv t, \; x^{n} \equiv r, \; x^{i} \;\; \mbox{where} 
\;\; i = 1, 2, \dots, (n-2) \right\} \,. \]
We think of $x^{i}$ as a parameterization of the local coordinates on an
$(n-2)$-sphere whose radius equals $\rho$, i.e. we take \( \rho^{2} = x_{i} x^{i} \), 
and consider the spacetime metric to be in the form \footnote{Note that the change of 
variable $\chi = \rho/(1 + \rho^{2}/4)$ transforms the metric (\ref{met}) to the 
following equivalent form:
\[ ds^{2} = - f^{2}(r) \, dt^{2} + u^{2}(r) \, dr^{2} + g^{2}(r) \, 
\left( \frac{d \chi^{2}}{1 - \chi^{2}} + \chi^{2} \, d \Omega^{2}_{n-3} \right) \,, \]
where $d\Omega^{2}_{n-3}$ denotes the metric on the unit $(n-3)$-sphere.}
\begin{equation}
ds^{2} = - f^{2}(r) \, dt^{2} + u^{2}(r) \, dr^{2} + g^{2}(r) \,
\sum_{i=1}^{n-2} \frac{dx_{i} \, dx^{i}}{(1 + \rho^{2}/4)^{2}} \;. \label{met}
\end{equation}
We choose the coframe 1-forms for the metric (\ref{met}) as
\begin{equation}
e^{0} = f(r) \, dt, \quad e^{n} = u(r) \, dr, \quad 
e^{i} = g(r) \, \frac{dx ^{i}}{(1 + \rho^{2}/4)} \,,
\;\; i = 1, 2, \dots, (n-2)  \,. \label{cof}
\end{equation}
Levi-Civita connection 1-forms follow easily from the first Cartan structure equations as
\begin{equation}
\omega^{0}\,_{i} = 0, \quad \omega^{i}\,_{j} = \frac{1}{2 g} (x^{i} e^{j} - x^{j} e^{i}), \quad
\omega^{0}\,_{n} = \frac{f^{\prime}}{f u} \, e^{0}, \quad
\omega^{i}\,_{n} = \frac{g^{\prime}}{u g} \, e^{i}, \label{con1f}
\end{equation}
where prime denotes derivative with respect to $r$. The curvature 2-forms that follow from
these read
\begin{equation}
R^{0n} = B \, e^{0} \wedge e^{n}, \quad R^{ij} = A \, e^{i} \wedge e^{j}, \quad
R^{0i} = C \, e^{0} \wedge e^{i}, \quad R^{in} = G \, e^{n} \wedge e^{i}, \label{cur2f}
\end{equation}
where we have used 
\begin{equation}
A = \frac{1}{g^{2}} \Big( 1 - \big( \frac{g^{\prime}}{u} \big)^{2} \Big), \quad
B = -\frac{1}{f u} \Big( \frac{f^{\prime}}{u} \Big)^{\prime}, \quad
C = -\frac{f^{\prime} g^{\prime}}{u^{2} f g}, \quad
G = \frac{1}{ug} \Big( \frac{g^{\prime}}{u} \Big)^{\prime}. \label{abcg}
\end{equation}

As for the YM potential 1-form, we employ the ansatz
\begin{equation}
{\cal A} = \frac{1}{2} \, \Sigma_{ij} \, \frac{x^{i} dx^{j} - x^{j} dx^{i}}{(1 + \rho^{2}/4)} 
\,, \label{gpot}
\end{equation}
where the matrices $\Sigma_{ij}$ denote the generators of the gauge group $SO(n-2)$
in the fundamental representation. Specifically, we choose them as
\begin{equation}
\Sigma_{ij}^{\alpha\beta} = \delta_{i}^{\alpha} \, \delta_{j}^{\beta} 
- \delta_{j}^{\alpha} \, \delta_{i}^{\beta} \,, \label{sig}
\end{equation}
with \( 1 \leq \alpha < \beta \leq n-2 \). This choice leads to the $\mathfrak{so}(n-2)$ 
commutation relations
\begin{equation}
[ \Sigma_{ij}, \Sigma_{k \ell} ] = 2 \, ( \delta_{\ell [ i} \, \Sigma_{j]k}
- \delta_{k [ i} \, \Sigma_{j] \ell} ) \,, \label{sigal}
\end{equation}
so that one obtains via (\ref{ym2f}) the YM 2-form field strength to be
\begin{equation}
{\cal F} = \frac{1}{2} \, \Sigma_{ij} \, \frac{dx^{i} \wedge dx^{j}}{(1 + \rho^{2}/4)^{2}}
= \frac{1}{2 g^{2}} \, \Sigma_{ij} \, e^{i} \wedge e^{j}  \,. \label{fcal}
\end{equation}
It is not hard to show that ${\cal F}$ satisfies (\ref{bian}) and (\ref{ymeq}) thanks to
(\ref{sigal}). Our choice (\ref{sig}) also leads to
\begin{equation} 
\mbox{Tr} \, (\Sigma_{ik} \, \Sigma_{kj}) = 2(n-3) \, \delta_{ij} \quad \mbox{and}
\quad \sum_{i<j} \, \mbox{Tr} \, (\Sigma_{ij} \, \Sigma_{ij}) = -(n-2)(n-3) 
\,. \label{sigcon}
\end{equation}

Note that (\ref{gpot}), and thus (\ref{fcal}), satisfy the flat space YM equations as 
well. Therefore, before moving onto the gravitational field equations, we want to make 
a digression and consider Yang's problem reviewed in section \ref{mono} for $n=6$ with 
gravitation still turned off. This time we want to replace the $SU(2)$ gauge group by 
$SO(4) \simeq (SU(2) \times SU(2))/\mathbb{Z}_{2}$. Following the discussion above, the 
corresponding bundle is $P(S^{4},SO(4))$ and the relevant homotopy group 
$\pi_{3}(SO(4))$ equals $\mathbb{Z} \oplus \mathbb{Z}$, therefore one may be inclined to think 
that \emph{if} there are solutions, their charges should be labelled by two independent 
integers. However, this is not the whole story since the gauge fields do have to satisfy 
the Euclidean YM equations as well. Using the 2-form field strength (\ref{fcal}) (with $g=1$) 
for $n=6$, if one naively calculates the charge as before using the analogous expression, one 
immediately finds
\[ \int_{S^{4}} \, \mbox{Tr} \, ({\cal F} \wedge {\cal F}) = \int_{S^{4}} ch_{2}({\cal F}) =  0 \,. \]
Nevertheless, one is saved by the topological quantity that takes the place of the charge 
which turns out to be the Euler characteristic given by \cite{hp, naka}
\[ \chi(S^{4}) = \frac{1}{32 \pi^{2}} \, \int_{S^{4}} \, \epsilon_{\alpha\beta\gamma\delta} \, 
{\cal F}^{\alpha\beta} \wedge {\cal F}^{\gamma\delta} = \frac{1}{128 \pi^{2}} \, 
\int_{S^{4}} \, \epsilon_{\alpha\beta\gamma\delta} \, \Sigma_{ij}^{\alpha\beta} \,
\Sigma_{kl}^{\gamma\delta} \, \epsilon^{ijkl} \, \hat{\ast} 1_{(4)} = 2 \,, \]
where $\hat{\ast} 1_{(4)}$ denotes the volume element of the $4$-sphere. We remark that 
for $n \geq 6$ and $n$ even with the gauge group $SO(n-2)$, a similar argument goes through 
analogously. Namely
\[ \int_{S^{n-2}} \, \mbox{Tr} \, {\cal F}^{(n-2)/2} = \int_{S^{n-2}} ch_{(n-2)/2}({\cal F}) =  0 \,, \]
and the Euler characteristic reads \( \chi(S^{n-2}) = 2 \). In fact, for generic $n$
\[ \chi(S^{n-2}) = \left\{
\begin{array}{ll}
0 \,, & n \; \mbox{is odd} \\
2 \,, & n \; \mbox{is even}
\end{array} \right. \,, \]
and since the Euler characteristic vanishes for any odd-dimensional manifold, one is urged to set
$n$ even. Thus from now on, we always take $n \geq 6$ and even. The solutions thus obtained 
are what we mean by flat-space Yang monopoles in higher (even) dimensions.

Finally, turning on gravity, the use of (\ref{sigcon}) in (\ref{eingb}) lead to the following 
system of coupled ordinary differential equations:
\begin{eqnarray}
B + (n-3) \Big( \frac{n-4}{2} A + C - G \Big) & = & \frac{(n-2)(n-3)}{4 \sigma^{2} g^{4}} - \Lambda \nonumber \\
 & & - \tilde{\gamma} \Big( A B - 2 C G + A (n-5) \big( C - G + \frac{n-6}{4} A \big) \Big) \,, \label{gb1} \\
(n-2) \Big( \frac{n-3}{2} A + C \Big) & = & - \frac{(n-2)(n-3)}{4 \sigma^{2} g^{4}} - \Lambda
- (n-2) \tilde{\gamma} A \Big( \frac{n-5}{4} A + C \Big) \,, \label{gb2} \\
(n-2) \Big( \frac{n-3}{2} A - G \Big) & = & - \frac{(n-2)(n-3)}{4 \sigma^{2} g^{4}} - \Lambda
- (n-2) \tilde{\gamma} A \Big( \frac{n-5}{4} A - G \Big) \,, \label{gb3}
\end{eqnarray}
where we have defined and used $\tilde{\gamma} = (n-3)(n-4) \gamma$.

\section{\label{solns} The solutions and their properties}

Setting \( u(r) =1/f(r) \) in (\ref{abcg}), one finds that (\ref{gb2}) and (\ref{gb3}) yield
\( g^{\prime\prime} = 0 \), and this leads to two independent cases: Either \textbf{i)} 
\( g(r) = g_{0} = \) constant or \textbf{ii)} \( g(r) = r \). It follows
that (\ref{gb1}), (\ref{gb2}) and (\ref{gb3}) admit two classes of solutions
corresponding to each case:

\textbf{i)} The first case leads to a cylindrical metric
\begin{equation}
ds^{2} = - f^{2}(r) \, dt^{2} + \frac{dr^{2}}{f^{2}(r)} + g_{0}^{2} \,
\sum_{i=1}^{n-2} \frac{dx_{i} \, dx^{i}}{(1 + \rho^{2}/4)^{2}} \,, \label{2ndsol}
\end{equation}
where 
\( f^{2}(r) = C_{0} \, r^{2} + C_{1} \, r + C_{2}  \,. \)
Here $C_1$ and $C_2$ are integration constants, and $C_{0}$ is given by
\[ C_{0} = - \frac{1}{g_{0}^{2} + \tilde{\gamma}} \, \left( \frac{(n-2)(n-3)}{2 \sigma^{2} g_{0}^{2}} 
+ \frac{n(n-3)}{4} + \frac{(n-5)\tilde{\gamma}}{g_{0}^{2}} \right) \,. \]
Note that the metric (\ref{2ndsol}) is conformally flat when $C_{0} g_{0}^{2} = 1$, which 
was also observed in \cite{halil}. We will not be interested in this solution.

\textbf{ii)} The second case is definitely more interesting and leads to the cosmological 
Einstein-GB Yang-monopole type solutions
\begin{equation}
ds^{2} = - f^{2}(r) \, dt^{2} + \frac{dr^{2}}{f^{2}(r)} + r^{2} \,
\sum_{i=1}^{n-2} \frac{dx_{i} \, dx^{i}}{(1 + \rho^{2}/4)^{2}} \,, \label{1stsol}
\end{equation}
where
\begin{equation}
f^{2}(r) = 1 + \frac{r^{2}}{\tilde{\gamma}} \left( 1 \pm
\sqrt{ 1 - 4 \tilde{\gamma} \left( \frac{\Lambda}{(n-2)(n-1)} - M r^{1-n}
+ \frac{(n-3)}{4 \sigma^{2} (n-5)} r^{-4} \right) } \, \right) \label{fdef}
\end{equation}
now. Here, as we will see, the constant $M$ is related to the gravitational mass 
of the solution.

Before we move on to studying the physical properties of this solution, we should note
that there is yet another, perhaps simpler, way of
obtaining the solutions (\ref{2ndsol}) and (\ref{1stsol}). It is based on inserting in
the action (\ref{act}) (and (\ref{lag})) the gauge fixed, static, spherically symmetric
metric (\ref{met}) with the corresponding YM field content calculated using (\ref{fcal})
and (\ref{sigcon}). This method was originally introduced by Weyl \cite{weyl} for obtaining
the exterior Schwarzschild solution of General Relativity, but was put on solid ground
much later in \cite{palais}. [See also \cite{dt} and \cite{dst} for some applications of 
this technique to various theories of gravitation.] The method considerably simplifies
the labor involved in obtaining the relevant field equations. Moreover, one can also use
it to show that the Birkhoff's theorem holds for the solution (\ref{2ndsol}): If the functions $f$ and
$u$ in the metric (\ref{met}) are also allowed to depend on the time coordinate $t$, the
Lagrangian density (\ref{lag}) turns out to be $t$-independent \cite{dt, df}. Thus
all spherically symmetric solutions are static in this model.

Let us look at various limits of this solution. For $\gamma \to 0$, we recover the
solutions presented in \cite{gib} by choosing the $-$ branch. When one takes 
$\Lambda = 0$ and $\sigma \to \infty$ in (\ref{fdef}), one recovers the external 
solutions of the Einstein-GB theory given in \cite{bode}. The branching of the solutions 
with either a Schwarzschild \( f^{2}(r) = 1 - 2 M r^{3-n} \) or a Schwarzschild-AdS 
\( f^{2}(r) = 1 + 2 M r^{3-n} + 2 r^{2}/\tilde{\gamma} \) type of asymptotics is 
recovered. For both sign choices in $f^{2}(r)$, the gravitational energy is found to 
be (up to some normalizations) proportional to $M$ by employing the energy 
definition of \cite{dt1, dt2}.

Now we consider the singularity structure of our solution. It is clear that there
is a curvature singularity at $r=0$, which follows from 
\( R_{ab} \wedge \ast R^{ab} = {\cal O}(r^{1-n}) \ast 1 \). There is an event horizon 
at $r_{H}>0$ ($f^{2}(r_{H})=0$), depending on the choice of parameters. In the most 
general case, this is a complicated analysis, but can be carried out along the lines 
of \cite{tm}. For simplicity, we concentrate on $n=6$ (the case of the Yang monopole) 
with $\Lambda=0$ and $\gamma \neq 0$ \footnote{$\gamma = 0$ case was considered in 
\cite{gib}.}. For this choice the location of the event horizon is given by the roots 
of the equation
\[ r^{3} + 3 \Big( \frac{1}{2 \sigma^{2}} + \gamma \Big) - 2 M = 0 \,, \]
which always has a real root $r_{H}$ if
\[ \Big( \frac{1}{2 \sigma^{2}} + \gamma \Big)^{3} + M^{2} \geq 0 \,, \]
and moreover, that root is positive if $M>0$ and $\gamma>0$.

Let us now compute the mass of this solution. Given a background Killing vector 
$\bar{\xi}^{\mu}$, the corresponding conserved charges of the model (\ref{lag}) 
can be written as \cite{dt1, dt2}
\begin{equation} 
Q^{\mu}(\bar{\xi}) = \frac{1}{4 \, \Omega_{n-2}} \,
\sqrt{1-\frac{4 \, \Lambda \, \tilde{\gamma}}{(n-1)(n-2)}} \,
 \int d^{n-2} x \, \bar{\xi}_{\nu} \, {\cal G}^{\mu\nu}_{L} \,. \label{yuk}
\end{equation}
Note that all the information coming from the GB part is encoded in the coefficient. 
The correct background to work with is the spacetime (\ref{1stsol}) with $M=0$ in 
(\ref{fdef}), and of course, with the timelike Killing vector 
\( \bar{\xi}^{\mu} = (-1, 0, \dots, 0) \) again. For convenience we also choose 
the $-$ branch \footnote{One can also proceed with the $+$ branch, in which case 
the spacetime is asymptotically AdS.}, which is asymptotically flat. One then finds 
the total energy according to (\ref{yuk}) as
\[ E = \frac{1}{4 \Omega_{n-2}} \, \frac{2(n-2)M}{\sqrt{1 - \frac{4 \Lambda \tilde{\gamma}}{(n-1)(n-2)}}} 
\, \sqrt{1 - \frac{4 \Lambda \tilde{\gamma}}{(n-1)(n-2)}} \, \Omega_{n-2} = \frac{(n-2)M}{2} \,, \]
which is finite.

\section{\label{conc} Conclusions}

We have shown that, contrary to the claim in \cite{gib}, the Yang monopole defined in even 
dimensions has a finite mass once gravity is introduced. This has been achieved by employing 
the method developed in \cite{ad, dt1, dt2} for which a proper choice of background is 
essential. Specifically, we have shown that out of the three generic parameters $m, \mu$
and $\Lambda$ of the gravitating Yang monopole, the first one can be interpreted as a 
\emph{mass} once the remaining two are allowed to constitute the background. 

We have also extended the family of Yang-monopole type solutions by studying the cosmological 
Einstein-GB-YM theory in higher even dimensions. We have also shown that these solutions 
have black hole singularities and event horizons for a proper choice of parameters.

Throughout this work, our discussion has been relying on $SO(n-2)$ gauge theory and
on static spherically symmetric $n$-dimensional metrics. If one abandons spherical
symmetry, one ends up with quite a nontrivial task of solving highly complicated 
differential equations. For example there is no solution describing a \emph{rotating} 
Yang monopole. As for the case of the (cosmological) Einstein-GB theory, the problem
is even harder: Let alone a rotating Yang monopole, there are no known exact 
rotating black hole solutions. 

\begin{acknowledgments}
We would like to thank Y{\i}ld{\i}ray Ozan and Turgut {\"O}nder for useful
discussions. This work is partially supported by the Scientific and Technological 
Research Council of Turkey (T{\"U}B\.{I}TAK). B.T. is also partially supported by
the Turkish Academy of Sciences (T{\"U}BA) and by the T{\"U}B\.{I}TAK Kariyer 
Grant 104T177. 
\end{acknowledgments}


\begin{thebibliography}{99}

\bibitem{yang}
C.N. Yang,
  J. Math. Phys. {\bf 19}, 320 (1978).

\bibitem{zhang}
S.C. Zhang and J. Hu,
  Science {\bf 294}, 823 (2001)
  [arXiv:cond-mat/0110572].

\bibitem{gib}
G.W. Gibbons and P.K. Townsend,
  Class. Quant. Grav. {\bf 23}, 4873 (2006)
  [arXiv:hep-th/0604024].

\bibitem{hp}
Z. Horvath and L. Palla,
  Nucl. Phys. B {\bf 142}, 327 (1978).

\bibitem{rt}
E. Radu and D.H. Tchrakian,
  Phys. Rev. D {\bf 71}, 125013 (2005)
  [arXiv:hep-th/0502025].

\bibitem{khn}
H. Kihara, Y. Hosotani and M. Nitta,
  Phys. Rev. D {\bf 71}, 041701 (2005)
  [arXiv:hep-th/0408068].

\bibitem{eh}
T. Eguchi, P.B. Gilkey and A.J. Hanson,
  Phys. Rept. {\bf 66}, 213 (1980).

\bibitem{naka}
M. Nakahara,
  ``{\it Geometry, Topology and Physics}'', Bristol: IOP Publishing, (1996).

\bibitem{ad}
L.F. Abbott and S. Deser,
  Nucl. Phys. B {\bf 195}, 76 (1982).

\bibitem{dt1}
S. Deser and B. Tekin,
  Phys. Rev. Lett. {\bf 89}, 101101 (2002)
  [arXiv:hep-th/0205318].

\bibitem{dt2}
S. Deser and B. Tekin,
  Phys. Rev. D {\bf 67}, 084009 (2003)
  [arXiv:hep-th/0212292].

\bibitem{bm}
R. Bartnik and J. McKinnon,
  Phys. Rev. Lett. {\bf 61}, 141 (1988).

\bibitem{halil}
S.H. Mazharimousavi and M. Halilsoy,
  ``Black Holes in Einstein-Maxwell-Yang-Mills Theory and their Gauss-Bonnet Extensions,''
  arXiv:0801.2110 [gr-qc].

\bibitem{weyl}
H. Weyl, ``{\it Space-Time-Matter}'', New York: Dover, (1951).

\bibitem{palais}
R.S. Palais, 
  Comm. Math. Phys. {\bf 69}, 19 (1979).

\bibitem{dt}
S. Deser and B. Tekin,
  Class. Quant. Grav. {\bf 20}, 4877 (2003) 
  [arXiv:gr-qc/0306114].

\bibitem{dst}
S. Deser, {\" O}. Sar{\i}o{\u g}lu and B. Tekin,
  Gen. Rel. Grav. {\bf 40}, 1 (2008)
  [arXiv:0705.1669 [gr-qc]].

\bibitem{df}
S. Deser and J. Franklin,
  Am. J. Phys. {\bf 73}, 261 (2005)
  [arXiv:gr-qc/0408067].

\bibitem{bode}
D.G. Boulware and S. Deser,
  Phys. Rev. Lett. {\bf 55}, 2656 (1985).

\bibitem{tm}
T. Torii and H. Maeda,
  Phys. Rev. D {\bf 72}, 064007 (2005)
  [arXiv:hep-th/0504141].

\end{thebibliography}
\end{document}